# Hybrid Correlation holography with a single pixel detector

Rakesh Kumar Singh'*

*Applied and Adaptive Optics Laboratory, Department of Physics, Indian Institute of Space Science and Technology (IIST), Trivandrum, 695547, Kerala, India*

*Corresponding author: krakeshsingh@iist.ac.in*

**The correlation holography reconstructs 3D objects as a distribution of two-point correlation of the random field detected by two dimensional detector arrays. Here, we describe a hybrid method, a combination of optical and computational channels, to reconstruct the objects from only a single pixel detector. An experimental arrangement is proposed and as a first step to realize the technique, we have simulated the experimental model for both scalar and vectorial objects. Proposed technique affords background free imaging with 3D capability.**

Holography records and reconstructs three dimensional (3D) information and is useful for non-destructive imaging. Availability of high quality sensors and computational facility has revolutionized the area through Digital Holography (DH) where the recorded hologram is reconstructed digitally rather than using optical means. DH reconstructs complex information and performs digital depth focusing [1]. This is currently a ubiquitous diagnostic and metrological tool. However, when the hologram is obstructed by an optically rough surface or illuminated by a random field, the phase and amplitude of light coming from it becomes scrambled hiding any deterministic information of the object encoded into the hologram. Scrambling of light field is connected to random distortion of the wavefront and formation of laser speckle due to complex interference of randomly scattered light field [2]. Imaging from the random field has garnered immense interest throughout the last several years and several techniques have been proposed [2-23]. Among them averaging the fluctuating scattered light through, say, correlation is very popular and is employed to image an object which is otherwise obscured into the laser speckle.

Correlation techniques such as ghost imaging [9-12], diffraction [13, 14] and correlation holography [17-23] have drawn significant attention. Ghost imaging provides object information from measurement of photon coincidence or intensity correlations between two beams. Correlation measurement is implemented for output from two photodetectors: one with a high spatial resolution, say CCD, which measures the field that has not interacted with the object to be imaged, and a bucket detector that collects the field that has interacted with the object. This was first demonstrated using the correlation of the quantum entangled photons state generated by parametric down conversion (PDC). Ghost imaging was later demonstrated with classical incoherent (pseudothermal) light source in the same way as entangled beams [10]. Recently, computational ghost imaging is proposed that uses only a single bucket detector [11, 24, 25]. Experimental implementation of the computational ghost imaging and ghost diffraction using a single detector has also been demonstrated by Bromberg et al [26]. By projecting a series of known random patterns and measuring the backscattered intensity with the help of several single pixel detectors, 3D form of the object is also reconstructed [27]. In a separate development, utilizing the phase shifting a single-pixel digital ghost holography has been implemented [28].

On the other hand, correlation holography, comprising of coherence holography, vectorial coherence holography and a photon correlation holography, reconstructs a 3D object as a distribution of the two point spatial correlation function. The principle behind this method is based on the van Cittert-Zernike theorem, which connects the far field two point correlation to the random source structure by a Fourier relation. Coherence holography reconstructs the object as distribution of the complex correlation by illuminating the digital hologram with incoherent light. However, measurement of the complex correlation function requires field based interferometer [17]. Vectorial coherence holography reconstructs the polarized object as distribution of the two point correlation of orthogonal polarization components of the random field [19]. We have demonstrated this technique by experimentally detecting polarized random fields and evaluating two point complex correlation from an interferometrically recorded data [19]. In another significant development, a technique termed photon correlation holography (PCH) has been developed to reconstruct the DH from the random field and to recover the object from two point intensity correlation of the far field speckle recorded by a array of photodetectors (CCD) [20]. This is possible because the intensity covariance is connected to the modulus of the complex correlation function and has its origin in the Hanbury Brown-Twiss interferometer [29].

However, none of correlation holography techniques are implemented with a single pixel detector. In this letter, we proposed hybrid correlation holography technique using only a single pixel detector. This technique is able to digital reconstruct the 3D object by estimating the cross-covariance of the random intensity patterns of a single pixel detector and digitally propagated random intensity patterns at the far field planes. The random light fields reaching the single point detector, in the optical channel, come after interaction with the object. On the other hand, digital propagated random fields, in the digital channel, do not interact with the object. Here, hybrid refers combination of optical and computational channels.

Comparison of the photon correlation holography (PCH) and proposed technique is shown in Fig. 1. Consider the setup shown in Fig. 1(a) which briefly describes the PCH. The monochromatic laser beam coming out of the laser is spatial filtered (S) and collimated by lens L. The collimated beam propagates through the rotating ground glass and creates the speckle pattern. This speckle pattern illuminates the transparency T as shown by a star in the black background, and subsequently random light field propagates down to the camera plane.

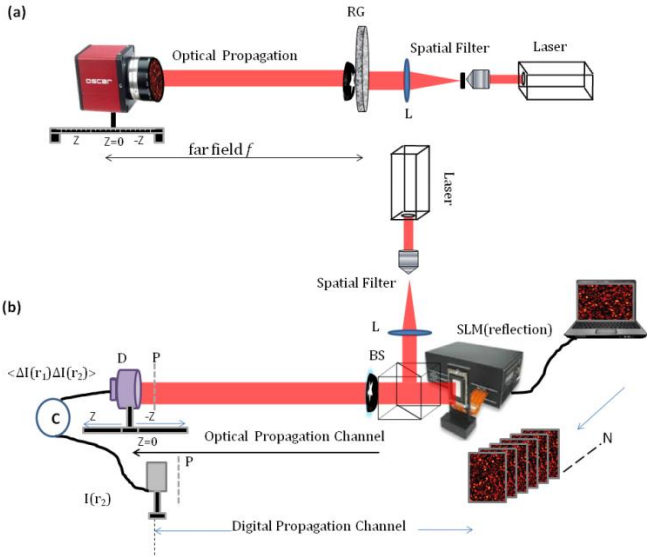

Fig. 1: (a) represents PCH setup, RG- rotating ground glass, L lens; (b). Hybrid correlation holography with single pixel detector setup: c- correlator, D- single pixel detector, BS- beam splitter, SLM- spatial light modulator, P- polarizer

The random intensity at CCD plane is represented as

$$I(\mathbf{u}) = |F\{T(\mathbf{r})\exp(i\phi(\mathbf{r}))\}|^2 \quad (1)$$

where $F$ is two dimensional Fourier transform, $T(\mathbf{r})$ represents transmittance (or reflectance) of the object information and $\phi(\mathbf{r})$ is the random phase introduced by the diffuser RG. In the above expression, $\mathbf{u}$ and $\mathbf{r}$ represents spatial coordinates at the detector and object planes respectively. Note that the intensity at different z can be obtained by using the proper propagation kernel. The cross-covariance of the intensity can be expressed as

$$\langle I(\mathbf{u})I(\mathbf{u}+\Delta\mathbf{u})\rangle = \langle \Delta I(\mathbf{u})\Delta I(\mathbf{u}+\Delta\mathbf{u})\rangle + \langle I(\mathbf{u})\rangle\langle I(\mathbf{u}+\Delta\mathbf{u})\rangle \quad (2)$$

where the angular bracket <.> denotes the ensemble average and random intensity variation from its mean intensity $\langle I(\mathbf{u})\rangle$ is $\Delta I(\mathbf{u}) = I(\mathbf{u}) - \langle I(\mathbf{u})\rangle$. For the Gaussian statistics, the intensity fluctuation can be expressed as

$$\langle \Delta I(\mathbf{u})\Delta I(\mathbf{u}+\Delta\mathbf{u})\rangle \propto |F[I_o(\mathbf{r})]|^2 \quad (3)$$

where $I_o(\mathbf{r}) = |T(\mathbf{r})|^2$ is intensity distribution at the random source plane. In deriving Eq. (3), delta correlation, i.e. $\langle e^{i[\phi(\mathbf{r}_2)-\phi(\mathbf{r}_1)]}\rangle = \delta(\mathbf{r}_2 - \mathbf{r}_1)$ and stationarity of the random field are exploited. Eq. (3) states that information encoded into transparency shapes the intensity co-variance. The transparency can be in the form of a hologram or aperture.

Let us now turn our attention to the hybrid correlation holography. In contrast to detector arrays and rotating ground glass (RG), our method applies a computer controlled spatial light modulator to introduce a sequence of random phases in the coherent beam in order to generate spatially incoherent source. A single pixel detector is used in the optical channel. The digitally stored random phase is also numerically propagated in the digital channel at z=0. Finally intensity correlation of the random fields coming from two arms, namely optical and digital, is evaluated. The proposed experimental arrangement is shown in Fig. 1 (b). The detailed theoretical explanation, procedure and results are discussed below.

A monochromatic laser beam, after spatial filtering assembly S and collimation lens L, enters into the beam splitter BS. The beam reflected from BS falls on the reflective type spatial light modulator (SLM), which introduces random phase to the incoming coherent light. The proposed SLM is in reflective mode, but same principle can also be applied for transmitive type SLM. The light after reflection from the SLM travels through the BS and stochastically illuminates transparency $T_i(\mathbf{r})$ for $i = x, y$ polarization components. To generalize the principle and to apply in the vectorial case, we establish the theory for coherent polarized object wherein dealing with two orthogonal polarization components is enough [21, 30, 31]. Optical channel contains a linear polarizer (for vectorial case) as shown by dotted line in front of the detectors. The polarized field vector is obtained by incorporating the Jones matrix $P(\theta)$ of the linear polarizer and represented as

$$\begin{pmatrix} E'_x(\mathbf{u}) \\ E'_y(\mathbf{u}) \end{pmatrix} = P(\theta) \begin{pmatrix} E_x(\mathbf{u}) \\ E_y(\mathbf{u}) \end{pmatrix} \quad (4)$$

where $P(\theta) = \begin{pmatrix} \cos^2\theta & \sin\theta\cos\theta \\ \sin\theta\cos\theta & \sin^2\theta \end{pmatrix}$ is the Jones matrix of the linear polarizer oriented at $\theta$ with respect to the horizontal (x) polarization state. The signal in the optical channel is proportional to the incident random intensity and represented as

$$I_o(\mathbf{u}) = |E'_{xo}(\mathbf{u})|^2 + |E'_{yo}(\mathbf{u})|^2 \quad (5)$$

Intensity for the digital channel can also be similarly expressed and intensity correlation between two channels is represented as

$$I_o(\mathbf{u}_1)I_c(\mathbf{u}_2) = \begin{bmatrix}(a^2+b^2)|E_{xo}(\mathbf{u}_1)|^2 + (b^2+c^2)|E_{yo}(\mathbf{u}_1)|^2 + (ab+bc)E_{xo}^*(\mathbf{u}_1)E_{yo}(\mathbf{u}_1) \\ (ab+bc)E_{yo}^*(\mathbf{u}_1)E_{xo}(\mathbf{u}_1)\end{bmatrix}$$
$$\begin{bmatrix}(a^2+b^2)|E_{xc}(\mathbf{u}_2)|^2 + (b^2+c^2)|E_{yc}(\mathbf{u}_2)|^2 + (ab+bc)E_{xc}^*(\mathbf{u}_2)E_{yc}(\mathbf{u}_2) \\ (ab+bc)E_{yc}^*(\mathbf{u}_2)E_{xc}(\mathbf{u}_2)\end{bmatrix}$$ (6)

where $a = \cos\theta$, $b = \sin\theta\cos\theta$, $c = \sin\theta$ and * stands for complex conjugate. Suffix o and c stands for optical and digital channels respectively. Due to Gaussian statistics, the intensity correlation function for $\theta = 0$ is

$$\langle I_o(\mathbf{u}_1)I_c(\mathbf{u}_2)\rangle = \langle E_{xo}^*(\mathbf{u}_1)E_{xo}(\mathbf{u}_1)\rangle \langle E_{xc}^*(\mathbf{u}_2)E_{xc}(\mathbf{u}_2)\rangle + \left|\langle E_{xc}^*(\mathbf{u}_1)E_{xo}(\mathbf{u}_2)\rangle\right|^2 \quad (7)$$

where <.> represents ensemble average. The intensity correlation for $\theta = \pi/2$ orientation is given as

$$\langle I_o(\mathbf{u}_1)I_c(\mathbf{u}_2)\rangle = \langle E_{yo}^*(\mathbf{u}_1)E_{yo}(\mathbf{u}_1)\rangle \langle E_{yc}^*(\mathbf{u}_2)E_{yc}(\mathbf{u}_2)\rangle + \left|\langle E_{yc}^*(\mathbf{u}_1)E_{yo}(\mathbf{u}_2)\rangle\right|^2 \quad (8)$$

Using angular spectrum method to propagate the deterministic random fields, the correlation of the random fields coming from two channels is written as

$$\langle E_{ic}^*(\mathbf{u},0)E_{io}(0,z)\rangle = \left\langle \iint T(\mathbf{r}_1)\exp[ik_z(\mathbf{r}_1)z]\exp\{i(\phi(\mathbf{r}_2)-\phi(\mathbf{r}_1))\}\exp[i2\pi\mathbf{u}\cdot\mathbf{r}_1]d\mathbf{r}_1d\mathbf{r}_2\right\rangle \quad (9)$$

The optical channel field $E_{io}(0,z)$ represents single point detector at transverse spatial location 0 and longitudinal position z. Using delta correlation feature of the random field, i.e. $\langle \exp[i(\phi(\mathbf{r}_2)-\phi(\mathbf{r}_1))]\rangle = \delta(\mathbf{r}_2 - \mathbf{r}_1)$, the correlation is

$$\langle E_{ic}^*(\mathbf{u},0)E_{io}(0,z)\rangle = \int T(\mathbf{r})\exp[ik_z(\mathbf{r})z]\exp[-i2\pi\mathbf{u}\cdot\mathbf{r}]d\mathbf{r} \quad (10)$$

where $z$ is propagation distance from the Fourier plane and $k_z = \frac{2\pi}{\lambda}\sqrt{1-\left(\frac{r}{f}\right)^2}$. Wavelength of light is $\lambda$ and $f$ is the Fourier transform distance. Using above relations and Gaussian statistics, the cross-covariance of the intensity becomes

$$\langle \Delta I_c(\mathbf{u},0)\Delta I_o(0,z)\rangle = \left|\int T(\mathbf{r})\exp[ik_z(\mathbf{r})z]\exp[-i2\pi\mathbf{u}\cdot\mathbf{r}]d\mathbf{r}\right|^2 \quad (11)$$

Eq. (11) states that 3D information of the object can be reconstructed. For demonstration purpose, let us consider transmittance function $T(\mathbf{r})$ as

$$T(\mathbf{r}) = \int\left[\int E(\mathbf{u},z)\exp\left[-i\frac{2\pi}{\lambda f}\mathbf{u}\cdot\mathbf{r}\right]d\mathbf{r}\right]\exp(ik_z z)dz \text{ and}$$

formed as explained in Fig. 2 for three different objects located at different z planes.

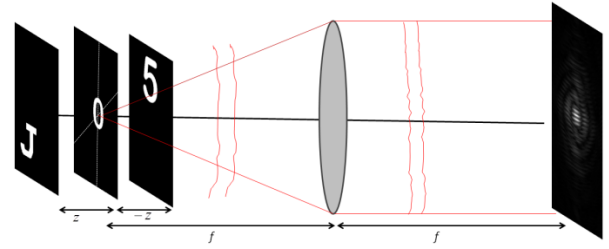

Fig. 2: Transparency encoded by 3D information of three objects located at three different z planes.

Implementation of the proposed technique is explained as follows. The SLM imposes a sequence of random phases $\phi_{nm}(M)$ to the incident light on nm pixel, where M represents random phases number. Random phases are independent and assumed to obey $\langle e^{i\phi_{nm}(M)}\rangle = 0$ and $\langle e^{i[\phi_{nm}(M_2)-\phi_{jk}(M_1)]}\rangle = \delta_{jn}\delta_{km}$. The coherent beam loaded with the random phase illuminates the transparency and field further propagates down to the single pixel detector. The random phase, displayed on the SLM, is also digitally propagated without the transparency for digital channel. The cross-covariance of the intensity is given as

$$\langle \Delta I_{io}(0,z)\Delta I_{ic}(\mathbf{u},0)\rangle = \sum_{n=1}^{M}\Delta I_{io}^n(0,z)\Delta I_{ic}^n(\mathbf{u},0) \quad (12)$$

In each realization of random pattern on the SLM, a random phase of size 400X400 is considered. The cross-covariance is obtained by correlating the variance of the speckle intensity from optical and digital arms and averaging over M different random patterns. Results of the cross-covariance are shown in Fig. 3. Figs. 3(a)-3(c) represent objects at three different z values for M = 4 00000. To reconstruct the object at different depths, the single pixel detector is scanned for z and corresponding intensity is correlated with digital channel intensity at z=0. Finally ensemble average of the cross covariance of the intensity is obtained. Fig. 3(a) shows reconstruction of the letter J at z=-200mm from the focal plane and other two letters are out of focus. Results for z=0 and z=100mm are shown in Fig. 3(b) and 3(c) respectively.

In order to quantitatively characterize the image reconstruction with variation of M, we table the reconstruction efficiency and visibility. Visibility is defined as the extent to which the reconstruction is distinguishable from the background noise. It can be measured as the ratio of the average image intensity level in the region corresponding to the signal area to the average background intensity level. The reconstruction efficiency $\eta$ is defined as the ratio of the measured power in the signal region of the image to the sum of this and the measured power in the 'background' region.

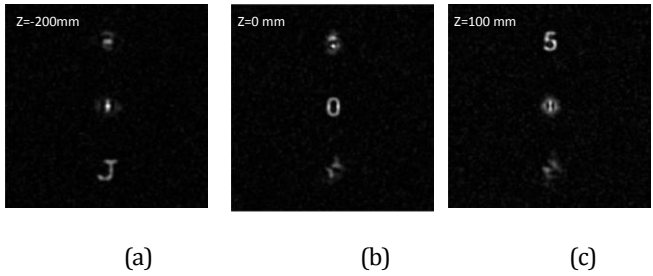

(a) (b) (c)

Fig. 3: Reconstructed image at different depth from focal plane (a) -200 (b) 0 (c) 100 in mm

Table- V represents visibility and R reconstruction efficiency

| M=5000 | M=50000 | M=100000 | M=400000 |
|---|---|---|---|
| V=3.48 R=0.7771 | V=9.760 R=0.9070 | V=11.9866 R=0.92299 | V=23.1649 R=0.958617 |

Reconstruction of the polarized coherent object from the correlation of the single point detector and digitally propagated random light field is shown in Fig. 4. Object transparency shown in Fig. 1b is, considered as composition of two orthogonally polarized holograms, illuminated by the random patterns. Display of the orthogonal polarized holograms and illumination by a common stochastic electromagnetic field can be experimentally realized by different geometries and one possible method is described by creating two replicas with a Sagnac interferometer [19]. This provides simultaneous illumination of the x and y polarized component holograms with a common stochastic field. Reconstruction results of the off-axis polarized holograms are shown in Fig. 4 (a) and 4(b) for θ=0 and 90. Reconstruction of the x and y polarized objects are obtained by performing the cross-covariance of the intensity of the corresponding polarized random fields from the optical and digital channels. The number of random fields M for ensemble averaging is 400000. Any error or displacement of the single point detector from the focal plane z=0 leads to the deformation in the reconstructed image. On axis power of the reconstructed image in Fig. 4(a) and 4(b) is suppressed in order to highlight reconstruction of the off-axis object as in the case of off-axis digital holography.

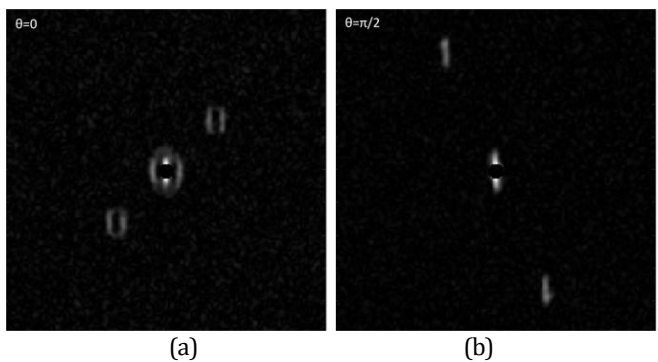

(a) (b)

Fig. 4: Reconstructed image from orthogonally polarized hologram at z=0 (a) x polarized (b) y polarized

In conclusion, a new technique to reconstruct the 3D information of the object from a single point detector is proposed. Based on the described theory, the computational experiment is carried out to test the proposed method and results are presented for scalar and vectorial cases. We have demonstrated reconstruction of the amplitude object and phase recovery is also possible by using a prior knowledge of a reference speckle. Proposed technique is expected to find wide range of applications in imaging and encryption.

**Funding.** Science Engineering Research Board-Department of Science and Technology (DST) India grant no EMR/2015/001613.

**Acknowledgment**. Author thanks Vinu RV for assistance in this work.